\documentclass[conference]{IEEEtran}
\IEEEoverridecommandlockouts

\usepackage{booktabs}
\usepackage{multirow}
\usepackage{graphicx}
\usepackage{subcaption}
\usepackage{caption}
\usepackage{float}
\usepackage{booktabs}
\usepackage{amsmath}
\usepackage{url}
\usepackage[margin=1in]{geometry}

\begin{document}

\title{Evaluating Japanese Dialect Robustness Across Speech and Text-based Large Language Models
}
\author{
\IEEEauthorblockN{
Tomoya Mizumoto, Yusuke Fujita, Hao Shi, Lianbo Liu, Atsushi Kojima, Yui Sudo
}
\vspace{0.1cm}
\IEEEauthorblockA{
\textit{SB Intuitions, Tokyo, Japan} \\
tomoya.mizumoto@sbintuitions.co.jp
}
}
%\author{
%\IEEEauthorblockN{Tomoya Mizumoto}
%\IEEEauthorblockA{Double-blind Review \\
%email address or ORCID
%}
%\and
%\IEEEauthorblockN{Yusuke Fujita}
%\and
%\IEEEauthorblockN{Hao Shi}
%\and
%\IEEEauthorblockN{Lianbo Liu}
%\and
%\IEEEauthorblockN{Atsushi Kojima}
%\and
%\IEEEauthorblockN{Yui Sudo}

\maketitle

\begin{abstract}
 Dialogue systems based on large language models (LLMs) have advanced significantly in recent years. 
 However, dialectal variation remains a major challenge, particularly for systems that process spoken input. 
 LLM-based speech language models (SLMs), which integrate LLMs with speech processing components, show promise for spoken language tasks, yet their ability to comprehend dialects has not been sufficiently studied. 
 Moreover, it remains unclear how the dialectal understanding of the base LLM affects SLM performance.
 This study investigates the dialectal robustness of both LLMs and SLMs using Japanese dialects as a test case. 
 We define robustness as the ratio of performance on dialectal versus standard inputs, enabling fair comparisons. 
 Our experiments show that SLM robustness correlates with that of their text-based counterparts.
 Furthermore, training with dialectal data and fine-tuning the speech encoder each improves robustness in SLMs.

\end{abstract}

\begin{IEEEkeywords}
  large language model, speech language model, speech translation, dialect
\end{IEEEkeywords}
  
\section{Introduction}
Among the various forms of linguistic diversity, dialects represent the most salient form of variation within a single language. 
Although large language models (LLMs) have demonstrated remarkable performance across a wide range of natural language processing tasks, including translation, summarization, and dialogue generation~\cite{qwen2025qwen25technicalreport,grattafiori2024llama3herdmodels,deepseekai2024deepseekv3technicalreport,openai2024gpt4technicalreport,oanil2023palm2technicalreport}, their ability to effectively process dialectal input remains limited. 
Dialects differ substantially from the standard variety in vocabulary, grammar, pronunciation, and pragmatics. 
Recent studies show that LLMs tend to underperform on dialectal input~\cite{lin2025languagegapsevaluatingdialect, limkonchotiwat2025assessingthaidialectperformance,ondrejova-suppa-2024-llms}. 
This shortcoming is further magnified in speech-based applications, where the variability of spoken dialects poses compounded challenges for both speech recognition and natural language understanding.

% 画像の挿入
\begin{figure}[t]
  \centering
  \includegraphics[width=\linewidth]{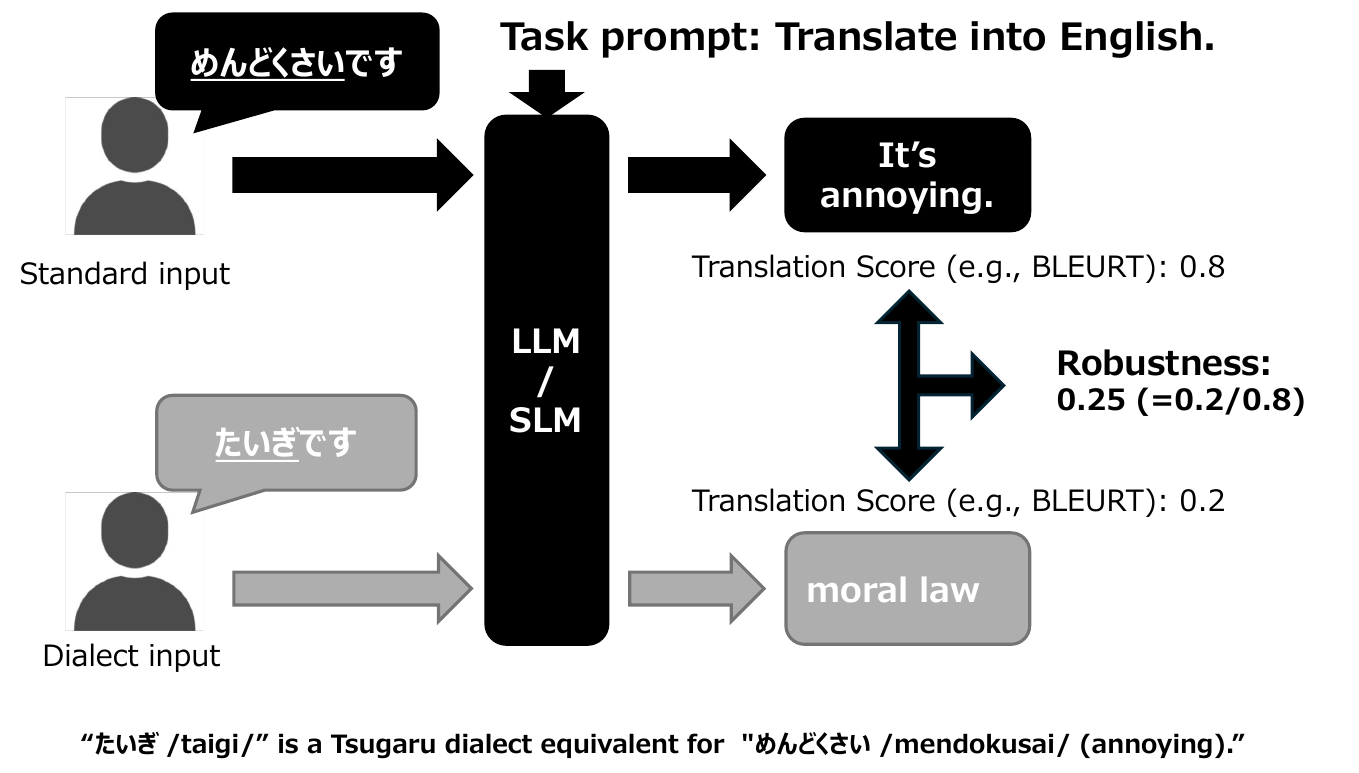}
  \vspace{-15pt}
  \caption{An example of robustness evaluation using a translation task. The standard and dialectal inputs convey the same meaning. Robustness is calculated based on the evaluation scores of the system outputs generated from standard and dialectal inputs.}
  \label{fig:robustness_example}
  \vspace{-10pt}
\end{figure}

In parallel, LLM-based speech language models (SLMs) have also made significant strides, with recent architectures moving beyond traditional speech processing pipelines by employing LLMs in end-to-end frameworks for tasks such as speech recognition, speech translation, and spoken dialogue~\cite{rubenstein2023audiopalm,chu2023qwenaudio,chu2024qwen2audiotechnicalreport}. 
However, it remains unclear to what extent these models can process dialectal speech, as systematic evaluations of regional variation are still limited.
This gap highlights the need for further investigation into the dialectal capabilities of LLM-driven speech systems and the development of methods to improve their inclusivity and linguistic adaptability.

 To address this gap, we conduct a comprehensive investigation of the dialectal robustness of both LLMs and SLMs. 
Evaluating how well a model handles dialectal input requires more than measuring absolute performance; it also demands consideration of factors such as model strength and evaluation difficulty.
We adopt a robustness-based evaluation approach that captures how reliably a model performs when the input shifts from standard language to dialectal forms. 
This evaluation perspective enables intuitive comparisons across models and conditions, and highlights each model’s capacity to generalize beyond standard language varieties. 
Fig.~\ref{fig:robustness_example} provides a simplified illustration of this concept.
Further details on the evaluation metrics and task setup are provided in Section~\ref{sec:eval_tasks}.
To enable consistent evaluation across modalities, we adopt a translation task from Japanese dialects to English as the primary setting for assessing dialectal comprehension.

To this end, our study addresses three key questions:
First, we examine whether the dialectal comprehension ability of a base LLM is preserved in its speech-based counterpart, thereby assessing the extent to which dialectal knowledge is transferred across modalities.
Second, we investigate whether incorporating dialectal data into SLM training improves performance on regional varieties, aiming to clarify the effect of expanded dialectal coverage.
Third, we evaluate the impact of fine-tuning the speech encoder on the dialectal robustness of SLMs, in order to isolate the contribution of the acoustic front-end to overall performance.
Through this analysis, we aim to shed light on the mechanisms underlying dialectal generalization in speech-capable LLMs and provide practical insights for building more inclusive, dialect-aware language technologies.
While our analysis focuses on Japanese dialects, we believe the findings have broader implications for multilingual and multi-dialectal speech systems.

\section{Related Work}
 Dialectal variation has long posed challenges for language technologies. 
In text-based natural language processing, studies have investigated dialect normalization~\cite{kuparinen-etal-2023-dialect}, multi-dialect machine translation~\cite{abe-etal-2018-multi}, and the construction of dialect corpora~\cite{xu-etal-2018-building}. 
In the speech domain, prior work has addressed dialect-aware automatic speech recognition (ASR)~\cite{miwa2023} and speech translation~\cite{paonessa-etal-2023-dialect,Samin2024}. 
These studies underscore the importance of dialect-specific data and model adaptation, particularly when handling unseen dialects.

 With the advent of LLMs, research has increasingly focused on their capacity to process dialectal input.
Some recent studies have evaluated LLMs on dialectal translation and common-sense reasoning~\cite{ondrejova-suppa-2024-llms}. 
In particular, Lin et al.~\cite{lin2025languagegapsevaluatingdialect} and Limkonchotiwat et al.~\cite{limkonchotiwat2025assessingthaidialectperformance} conducted evaluations directly comparing performance on standard versus dialectal inputs, reporting substantial degradation for dialects and underscoring the lack of robustness in current LLMs.

 While these studies have provided valuable insights, most focus on either text or speech in isolation and do not address whether dialectal competence generalizes across modalities.
Moreover, there has been limited investigation into the alignment between LLMs and their speech-based extensions (SLMs) in handling dialectal variation. 
Our study addresses this gap by systematically evaluating both LLMs and SLMs across 20 Japanese dialects, focusing on dialectal robustness and the conditions under which it can be improved

\section{Research Design} \label{sec:research_design}

 This study investigates the dialectal robustness of LLMs and SLMs using Japanese as a test case.
Specifically, we address three research questions:

\begin{itemize}
    \item \textbf{RQ1:} Is the dialectal comprehension ability of a base LLM preserved in its speech-capable variant (SLM)?
    \item \textbf{RQ2:} Does incorporating dialectal training data improve robustness to dialectal input?
    \item \textbf{RQ3:} Does fine-tuning the speech encoder enhance dialectal robustness in SLMs?
\end{itemize}

To answer these questions, we construct and evaluate models under multiple conditions, systematically varying the use of dialectal data during training and encoder fine-tuning while comparing LLM and SLM performance.

\subsection{Evaluation task and method} \label{sec:eval_tasks}
 We evaluate dialectal comprehension using a {\bf translation task from Japanese dialects to English}. 
This task was selected for its applicability to both text- and speech-based models, as well as its suitability for controlled and interpretable evaluation.
Although other tasks, such as ASR, dialect-to-standard translation, or dialogue response generation, may appear viable, they entail practical or conceptual limitations.
ASR is inherently inapplicable to text-input models, and dialect-to-standard translation is constrained by the scarcity of high-quality parallel data, making model training and evaluation unreliable. 
Response generation, by contrast, does not reliably capture comprehension: models can produce plausible replies without resolving dialect-specific expressions.
For example, given an input like “I ate xxx today” (where “xxx” is a dialectal expression), a model might respond with a generic utterance such as “Was it good?” even without recognizing the dialectal term.
In contrast, translation into English requires the model to interpret the full meaning of dialectal inputs at the utterance level. 
It demands disambiguation of regional lexical and syntactic variations, and the availability of gold-standard references enables objective comparison via automated metrics. 
Thus, translation provides a principled and scalable framework for evaluating dialectal robustness across both modalities.

To assess model quality, we employ standard automatic metrics widely used in machine translation.
Specifically, we use BLEU~\cite{Papineni-2002-bleu} and  BLEURT~\cite{sellam-etal-2020-bleurt} scores. 
BLEU evaluates surface-level similarity via n-gram overlap, whereas BLEURT captures semantic similarity by leveraging pretrained language models.
All metrics are computed against English reference translations.

Evaluating dialectal generalization requires not only measuring absolute performance but also assessing the consistency of performance across different language varieties.
Previous studies have quantified performance differences between standard and dialectal inputs using absolute score gaps~\cite{lin2025languagegapsevaluatingdialect,limkonchotiwat2025assessingthaidialectperformance}.
While informative, such gaps are sensitive to baseline performance and complicate comparisons across models, tasks, or datasets.
To address this, we explicitly define \textit{dialectal robustness} as the ratio of performance on dialectal inputs to that on standard inputs:
\begin{equation}
    \setlength{\abovedisplayskip}{2pt}
    \setlength{\belowdisplayskip}{2pt}
    \mathrm{Robustness} = S_{\text{dialect}}/S_{\text{standard}},
\end{equation}
where \( S_{\text{dialect}} \) and \( S_{\text{standard}} \) denote the evaluation scores (e.g., BLEU or BLEURT) obtained when the model is given dialectal and standard language inputs, respectively.
This normalized formulation enables interpretable, model-agnostic comparisons and clearly reflects a model’s ability to generalize beyond standard language. A robustness score near 1 indicates high dialect invariance, whereas lower values reflect greater sensitivity to dialectal variation. A schematic illustration of this framework is shown in Fig.~\ref{fig:robustness_example}.

\subsection{Model architecture}
For this study, we developed two model types: a text-to-text translation model based on an LLM, and a speech-enabled SLM capable of processing spoken input.
We provide an overview of each model below.

\begin{figure}[t]
  \centering
  \includegraphics[width=0.95\linewidth]{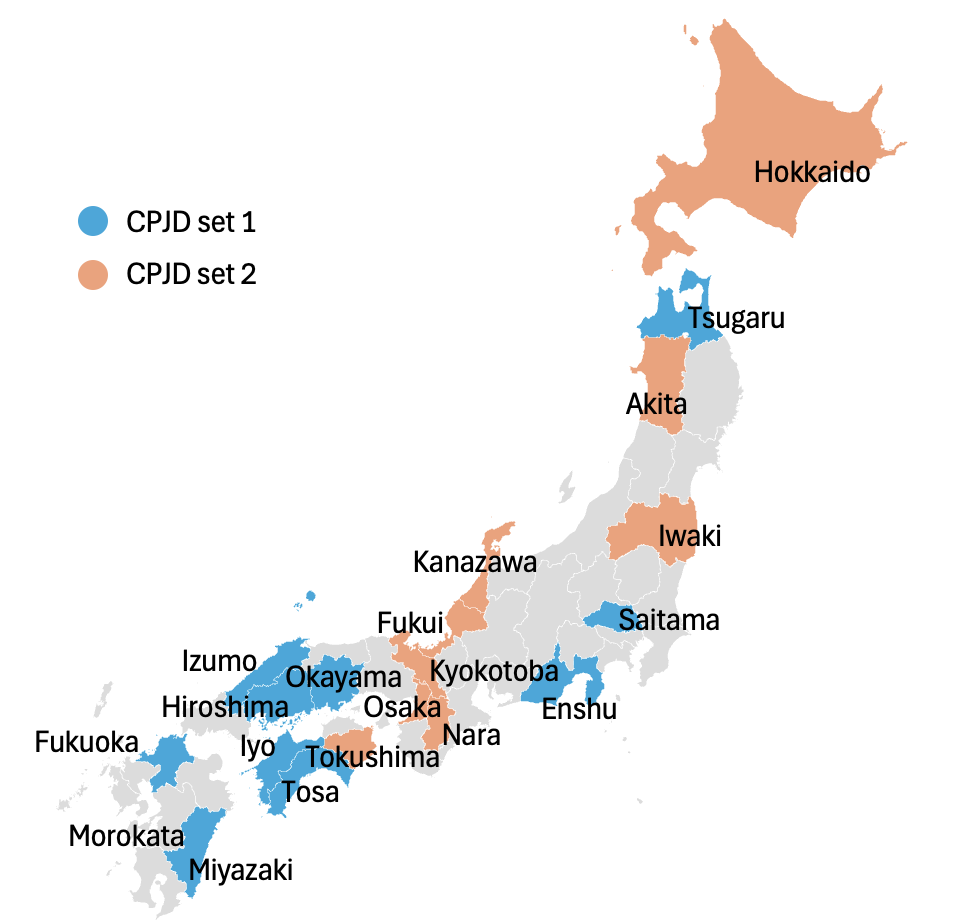}
  \caption{Geographic distribution of the 20 dialects in the CPJD corpus. Each region is colored according to the CPJD sentence set used by its speakers: Set 1 (blue) or Set 2 (orange).}
  \label{fig:map}
  \vspace{-10pt}
\end{figure}

\subsubsection{Text input model}
Instruction-tuned LLMs are generally capable of performing translation tasks without additional training when given appropriate prompts.
However, such models may produce extraneous output beyond the translation itself---for example, phrases like ``Here is the translation.''
This behavior can unfairly lower evaluation scores, particularly when using automatic metrics that assume clean translations.

To address this issue, we applied Low-Rank Adaptation (LoRA)~\cite{hu2022lora} to fine-tune the model so that it outputs only the translated text, without any additional explanatory phrases. 
This approach enables efficient fine-tuning and provides greater control over the model’s output, such as suppressing undesired prompt-following responses in translation tasks.

\subsubsection{Audio input model}
For speech-to-text translation, we develop models that take spoken utterances as input and generate translations as output.
The model is built by connecting a speech encoder to an LLM using an adapter module~\cite{hono-etal-2024-integrating,fathullah2024,Lakomkin2024}.

The adapter module adopts the architecture proposed by Hono et al.~\cite{hono-etal-2024-integrating}. 
It consists of a downsampling block followed by a linear projection layer, which reduces sequence length and aligns the feature dimensionality of speech representations with the LLM’s input embeddings.
Let \( \mathbf{H} \in \mathrm{R}^{T \times D} \) denote the hidden representation output from the speech encoder, where \( T \) is the sequence length and \( D \) is the feature dimension.  
The downsampling block applies two 1D convolutional layers along the time axis, reducing the sequence length by a factor of 4:
\begin{equation}
    \setlength{\abovedisplayskip}{2pt}
    \setlength{\belowdisplayskip}{2pt}
    \mathbf{H}' = \mathrm{Conv}_2(\mathrm{Conv}_1(\mathbf{H})) \in \mathrm{R}^{T/4 \times D'},
\end{equation}
where \( \mathrm{Conv}_1 \) and \( \mathrm{Conv}_2 \) are convolution layers with stride 2, kernel size \( k \), and output dimensionality \( D' \). 
Next, a linear projection aligns the downsampled features with the LLM's embedding dimension \( d_{\mathrm{LLM}} \):
\begin{equation}
    \setlength{\abovedisplayskip}{2pt}
    \setlength{\belowdisplayskip}{2pt}
    \mathbf{Z} = \mathbf{H}' \mathbf{W}_{\text{proj}} + \mathbf{b}_{\text{proj}} \in \mathrm{R}^{T/4 \times d_{\mathrm{LLM}}},
\end{equation}
where \( \mathbf{W}_{\text{proj}} \in \mathrm{R}^{D' \times d_{\mathrm{LLM}}} \) and \( \mathbf{b}_{\text{proj}} \in \mathrm{R}^{d_{\mathrm{LLM}}} \) are the trainable projection weights and bias, respectively.
The resulting representation $\mathbf{Z}$ is directly used as the input embedding sequence for the LLM.

\subsection{Data}~\label{sec:data}
\begin{table}[t]
  \centering
  \renewcommand{\arraystretch}{0.9}
  \caption{Overview of datasets used in the study.} 

  \vspace{-5pt}
  \label{table:data}
  \label{tab:dataset_overview}
  \begin{tabular}{llrll}
  \toprule
  \multirow{2}{*}{Dataset} & \multirow{2}{*}{Std/Dial} & Size & \multirow{2}{*}{Speaker} & English \\
  & & \multicolumn{1}{c}{(\#utt.)} & & Source\\
  \midrule
  \multicolumn{5}{c}{\textbf{Train}} \\
  \midrule
  Reazonspeech & standard & 2975935 & Human & Qwen2.5 \\
  CoVost              & standard & 1119 & Human & Human \\
  Speech BSD          & standard & 20000 & Human & Human \\
  CPJD1-train         & dialect  & 2640 & Human & Qwen2.5 \\
  CPJD2-train         & dialect  & 2401 & Human & Qwen2.5 \\
  \midrule
  \multicolumn{5}{c}{\textbf{Evaluation}} \\
  \midrule
  CPJD1-eval-dial & dialect  & 2640 & Human & GPT4o \\
  CPJD2-eval-dial & dialect  &  2401 & Human & GPT4o \\
  CPJD1-eval-std  & standard &  1250 & TTS   & GPT4o \\
  CPJD2-eval-std  & standard & 1250 & TTS   & GPT4o \\
  \bottomrule
  \end{tabular}
  \vspace{-10pt}
  \end{table}
\begin{table}[t]
  \centering
  \renewcommand{\arraystretch}{0.9}
  \caption{BLEU and BLEURT Scores and Robustness. Rbst represents robustness. Std and Dial refer to scores on standard and dialectal, respectively.}
  \label{table:result001}
  \begin{tabular}{llrrrrrr}
  \toprule \textbf{Eval} & \textbf{Input} & \multicolumn{3}{c}{\textbf{BLEU}} & \multicolumn{3}{c}{\textbf{BLEURT}}  \\
  \textbf{Data} & & Std & Dial & Rbst & Std & Dial & Rbst \\
  \midrule
  \multicolumn{8}{c}{\textbf{Llama}} \\
  \midrule
  \cmidrule(lr){3-5} \cmidrule(lr){6-8}
   
   \multirow{2}{*}{CPJD1}
   & Text & .274 & .230 & .839 & .715 & .678 & .948 \\
   & Audio & .295 & .206 & .698 & .694 & .635 & .915 \\
  \cmidrule(lr){1-8}
  \multirow{2}{*}{CPJD2}
    & Text & .280 & .203 & .725 & .720 & .644 & .895 \\
    & Audio & .297 & .190 & .640 & .699 & .617 & .883 \\
  \midrule
  \multicolumn{8}{c}{\textbf{LLMJP}} \\
  \midrule
  \multirow{2}{*}{CPJD1}
    & Text  & .301 & .258 & .857 & .730 & .700 & .959 \\
    & Audio & .206 & .160 & .777 & .653 & .606 & .929 \\
  \cmidrule(lr){1-8}
  \multirow{2}{*}{CPJD2}
    & Text  & .303 & .233 & .769 & .735 & .668 & .910 \\
    & Audio & .204 & .159 & .779 & .652 & .594 & .911 \\
  \midrule
  \multicolumn{8}{c}{\textbf{Sarashina}} \\
  \midrule
  \multirow{2}{*}{CPJD1}
    & Text  & .319 & .268 & .840 & .731 & .703 & .962 \\
    & Audio & .276 & .209 & .757 & .687 & .639 & .930 \\
  \cmidrule(lr){1-8}
  \multirow{2}{*}{CPJD2}
    & Text  & .294 & .238 & .810 & .733 & .672 & .917 \\
    & Audio & .270 & .196 & .726 & .693 & .626 & .904 \\
  \midrule
  \multicolumn{8}{c}{\textbf{Swallow}} \\
  \midrule
  \multirow{2}{*}{CPJD1}
    & Text  & .322 & .272 & .845 & .734 & .704 & .959 \\
    & Audio & .294 & .226 & .769 & .691 & .644 & .932 \\
    \cmidrule(lr){1-8}
  \multirow{2}{*}{CPJD2}
    & Text  & .330 & .255 & .773 & .748 & .675 & .903 \\
    & Audio & .314 & .223 & .710 & .700 & .631 & .902 \\
  \bottomrule
  \end{tabular}
  \vspace{-10pt}
  \end{table}
  
\vspace{-10pt}
%This section provides an overview of the data used in this study.
We employ two data types: general speech transcription data for training the SLM, and dialectal speech data for training and evaluating its ability to handle dialectal variation.

For general-purpose speech data, we use the ReazonSpeech v2.0 corpus~\cite{reazonspeech}, Speech BSD~\cite{shimizu-etal-2023-towards}, and CoVoST2~\cite{wang2020covost}.
The ReazonSpeech corpus consists of paired broadcast speech and subtitles from television programs.
Since using the entire dataset would be computationally expensive, we use a subset consisting of approximately 5,000 hours of audio and 3 million utterances.
Both Speech BSD and CoVoST provide speech data with corresponding transcriptions and translations.

For dialectal data, we use the Crowdsourced Parallel Speech Corpus of Japanese Dialects (CPJD)~\cite{takamichi-saruwatari-2018-cpjd}, which provides dialectal speech paired with transcriptions in both dialect and standard Japanese, but does not include standard-Japanese speech recordings.
Several other Japanese dialect corpora exist, but their utility is limited for different reasons: the Japanese Multi-Dialect Corpus for Speech Synthesis (JMD)\footnote{\url{https://sites.google.com/site/shinnosuketakamichi/research-topics/jmd_corpus}} lacks standard‑Japanese transcriptions, whereas the Corpus of Japanese Dialects (COJADS)\footnote{\url{https://www2.ninjal.ac.jp/cojads/index.html}} is not publicly available.

The CPJD corpus provides two sets of standard Japanese utterances as its base. 
Each speaker was assigned one of the two standard Japanese sentence sets and asked to read it aloud in their own dialect. 
This procedure yielded paired data between standard Japanese text and dialectal speech (with corresponding transcriptions), enabling both cross-variety and cross-modality comparisons.
The corpus covers 20 regional dialects, each with approximately 150 to 250 utterances. 
An exception is the Nara dialect, which includes 499 utterances from two speakers. 
Fig.~\ref{fig:map} illustrates the geographic distribution of dialects grouped by sentence set. 
The dialects are divided into two subsets, each aligned with one of the two standard utterance sets. 
However, the distribution is not fully balanced---some regions appear only in one subset. 

 Neither the ReazonSpeech dataset nor the CPJD corpus includes English translations.
Moreover, while CPJD provides standard Japanese text as a reference, it does not include corresponding audio recordings of these utterances.
To address the lack of translations and recordings, we generated English translations for both datasets using an LLM, and synthesized speech for the standard Japanese sentences in CPJD using a text-to-speech (TTS) system.
Given the large number of translations required for training, we employed the Qwen2.5-32B-Instruct model, which offers a favorable balance between translation quality and computational cost.
This approach was applied to both ReazonSpeech and CPJD.
In the case of CPJD, although both standard Japanese and dialectal utterances are available, we used the standard Japanese text as input for translation.
For evaluation, we used GPT-4o to generate higher-quality English translations of the standard Japanese portion of CPJD.
Table~\ref{table:data} summarizes the datasets used in this study, including their standard/dialectal status, size, speaker type, and the source of English translations.

\begin{figure*}[h]
  \centering
  \begin{subfigure}[t]{0.45\textwidth}
    \centering
    \includegraphics[width=\linewidth]{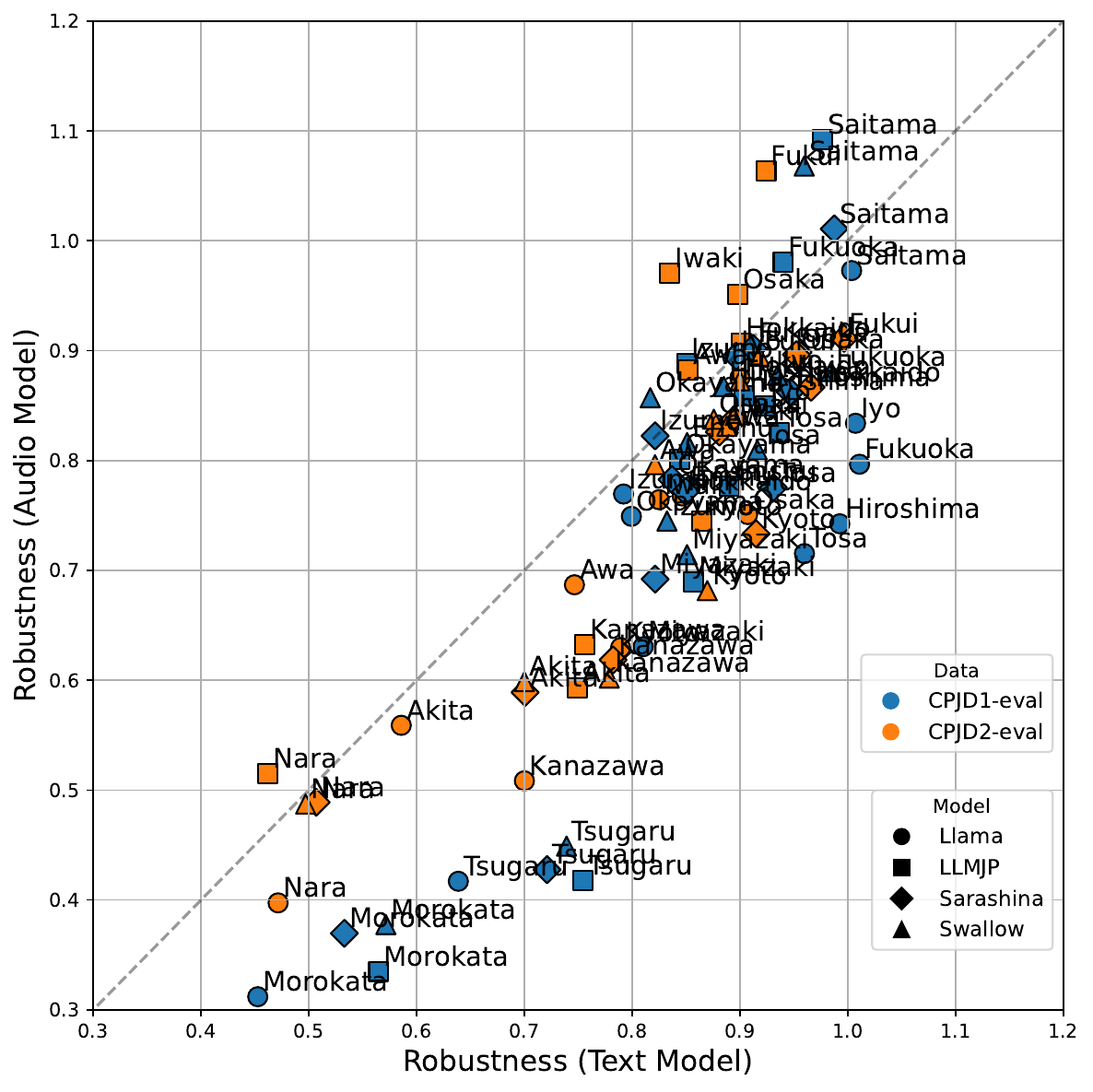}
    \vspace{-15pt}
    \caption{BLEU Robustness}
  \end{subfigure}
  \hfill
  \begin{subfigure}[t]{0.45\textwidth}
    \centering
    \includegraphics[width=\linewidth]{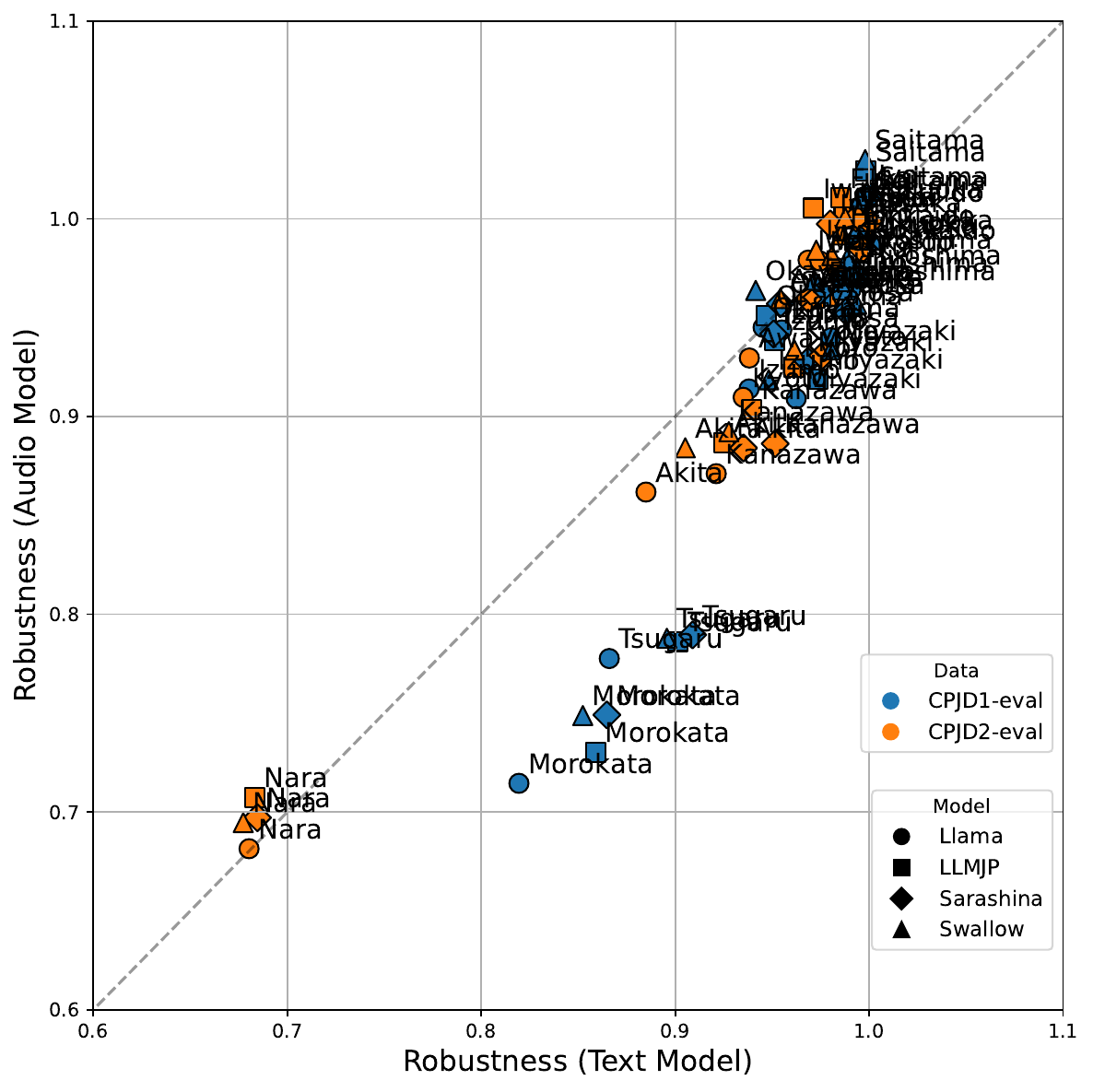}
        \vspace{-15pt}
    \caption{BLEURT Robustness}
  \end{subfigure}
  \vspace{-5pt}
  \caption{Robustness comparison between text and audio models. The x-axis and y-axis represent robustness scores for text and audio models, respectively. Shape indicates the model (Llama, LLMJP, Sarashina, or Swallow), and color denotes the evaluation split (CPJD1 or CPJD2).
  The dashed diagonal line represents equal robustness across modalities; points below the line indicate reduced robustness in the audio model.} 
  \vspace{-15pt}\label{fig:result002}
\end{figure*}

\section{Experiments}

The experiments are structured around the three research questions introduced in Section~\ref{sec:research_design}.

\subsection{Experimental settings}
 We first provide an overview of the LLMs and speech encoder used in this study.
As LLMs, we used Llama-3.1-8B-Instruct\footnote{\url{https://huggingface.co/meta-llama/Llama-3.1-8B-Instruct}} (Llama), which was trained primarily on English data, and three models designed for Japanese language: Llama-3.1-Swallow-8B-Instruct-v0.3\footnote{\url{https://huggingface.co/tokyotech-llm/Llama-3.1-Swallow-8B-Instruct-v0.3}}~\cite{Okazaki:COLM2024,Fujii:COLM2024} (Swallow), llm-jp-3-7.2b-instruct3\footnote{\url{https://huggingface.co/llm-jp/llm-jp-3-7.2b-instruct3}} (LLMJP), and sarashina2.2-3b-instruct-v0.1\footnote{\url{https://huggingface.co/sbintuitions/sarashina2.2-3b-instruct-v0.1}} (Sarashina).
For the speech encoder, we used only the encoder component of Whisper-Large-V3~\cite{radford2022whisper}.

In our experiments, the weights of the underlying LLM and speech encoder are fixed unless otherwise specified.
For RQ1 and RQ2, the weights of both the LLM and the speech encoder are kept frozen, while only the adapter modules and the LoRA parameters for text-based models are trained.
For RQ3, we investigate the impact of speech representation learning by fine-tuning the speech encoder while keeping the LLM fixed.

For training the SLMs, we use the AdamW optimizer with a cosine learning rate schedule decaying from \(1 \times 10^{-4}\) to \(1 \times 10^{-6}\), and apply a weight decay of 0.01.
Models are trained for 25{,}000 steps with 500 warmup steps, using 16 A100 GPUs (80 GB), a batch size of 8, and gradient accumulation over 4 steps. 
For experiments involving speech encoder fine-tuning, we extend the training to 50{,}000 steps with a reduced batch size of 4 and double the number of GPUs (32 A100s).
For text-based experiments, LoRA fine-tuning is performed on 8 A100 GPUs with 3{,}000 training steps and 100 warmup steps, using the same optimizer and learning rate configuration.

In the experiment for RQ1, models were trained on ReazonSpeech, CoVoST, and Speech BSD. 
For the text-based model, we applied LoRA fine-tuning using a subset of approximately 62,000 utterances from ReazonSpeech in order to mitigate overfitting risks associated with the large corpus size.
For RQ2 and RQ3, training was performed on the CPJD corpus. 
It is divided into two subsets, CPJD1 and CPJD2, corresponding to the sentence sets described in Section~\ref{sec:data}.
To avoid any overlap between training and evaluation data, we trained the model on one subset while evaluating on the other. 
Specifically, when evaluating on CPJD1, the model was trained on ReazonSpeech and CPJD2, and vice versa.
Previous work has shown that geographically proximate regions tend to exhibit similar dialectal features~\cite{abe-etal-2018-multi}.
In this split design, not only the target dialect but also neighboring dialects may be excluded from training.
We analyze the effect of such exclusions in our evaluation.

\subsection{RQ1: Do SLMs retain the dialectal strengths of their base LLMs?}
Table~\ref{table:result001} presents the overall results across all models.
All models exhibit consistently lower scores for dialectal input compared to standard Japanese. 
This performance drop is even more pronounced when the input is provided in spoken form.
These results highlight the persistent difficulty of handling dialectal input.
With the exception of standard Japanese input with Llama, speech input generally yields lower BLEU scores than text input.
One possible explanation is that the base Llama model had limited exposure to certain Japanese tokens; however, through training on speech data, it may have acquired associations between these tokens and their appropriate lexical representations.

\begin{figure}[h]
  \centering
  \includegraphics[width=0.45\textwidth]{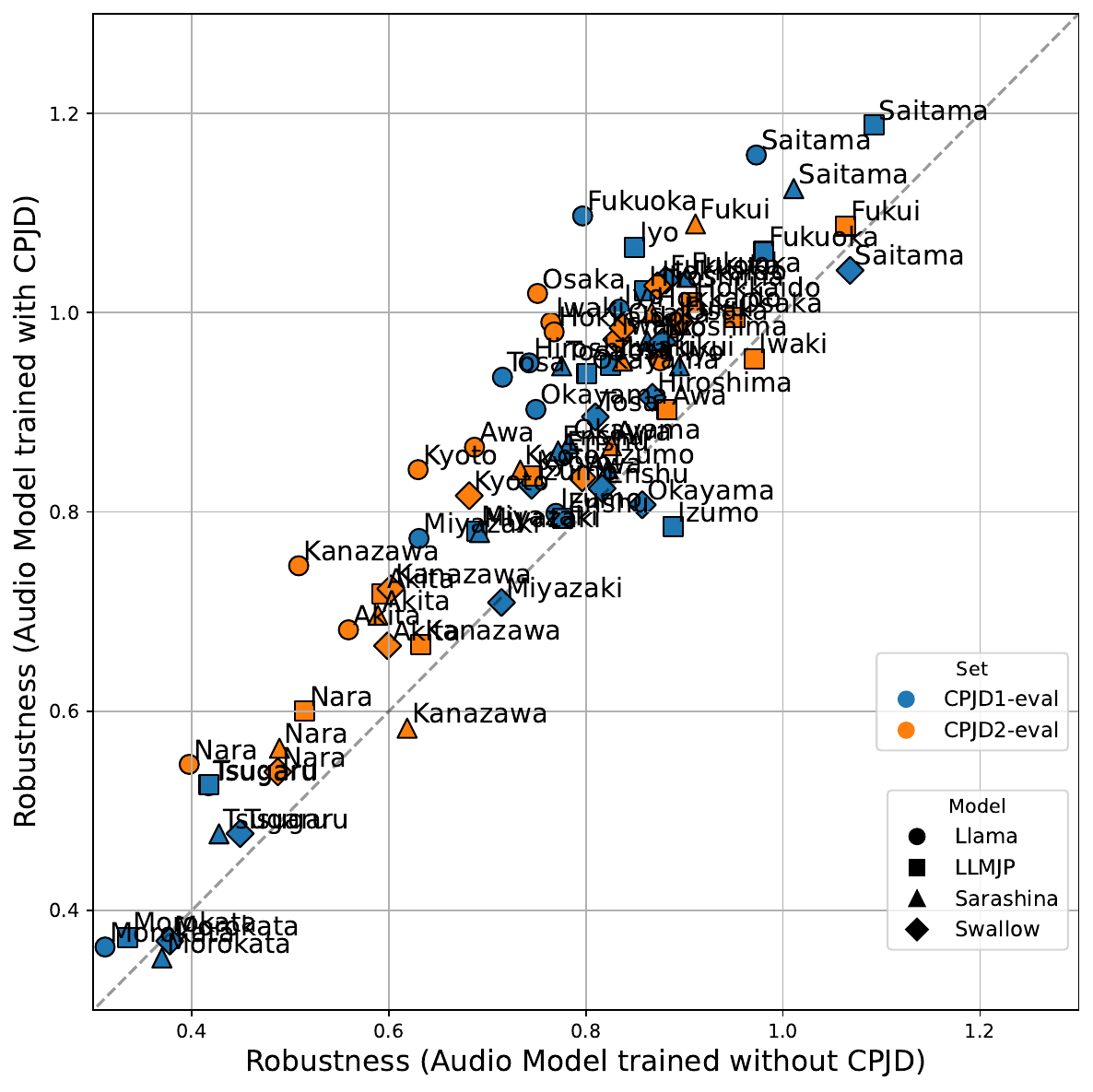}
  \vspace{-5pt}
  \caption{Robustness comparison between audio models trained with and without CPJD  (BLEU-based). The x-axis and y-axis represent dialectal robustness scores from models trained without and with CPJD, respectively. Points above the diagonal indicate improved robustness with CPJD training.} 
\vspace{-15pt}
\label{fig:result004}
\end{figure}
\begin{figure}[h]
  \centering
  \includegraphics[width=0.45\textwidth]{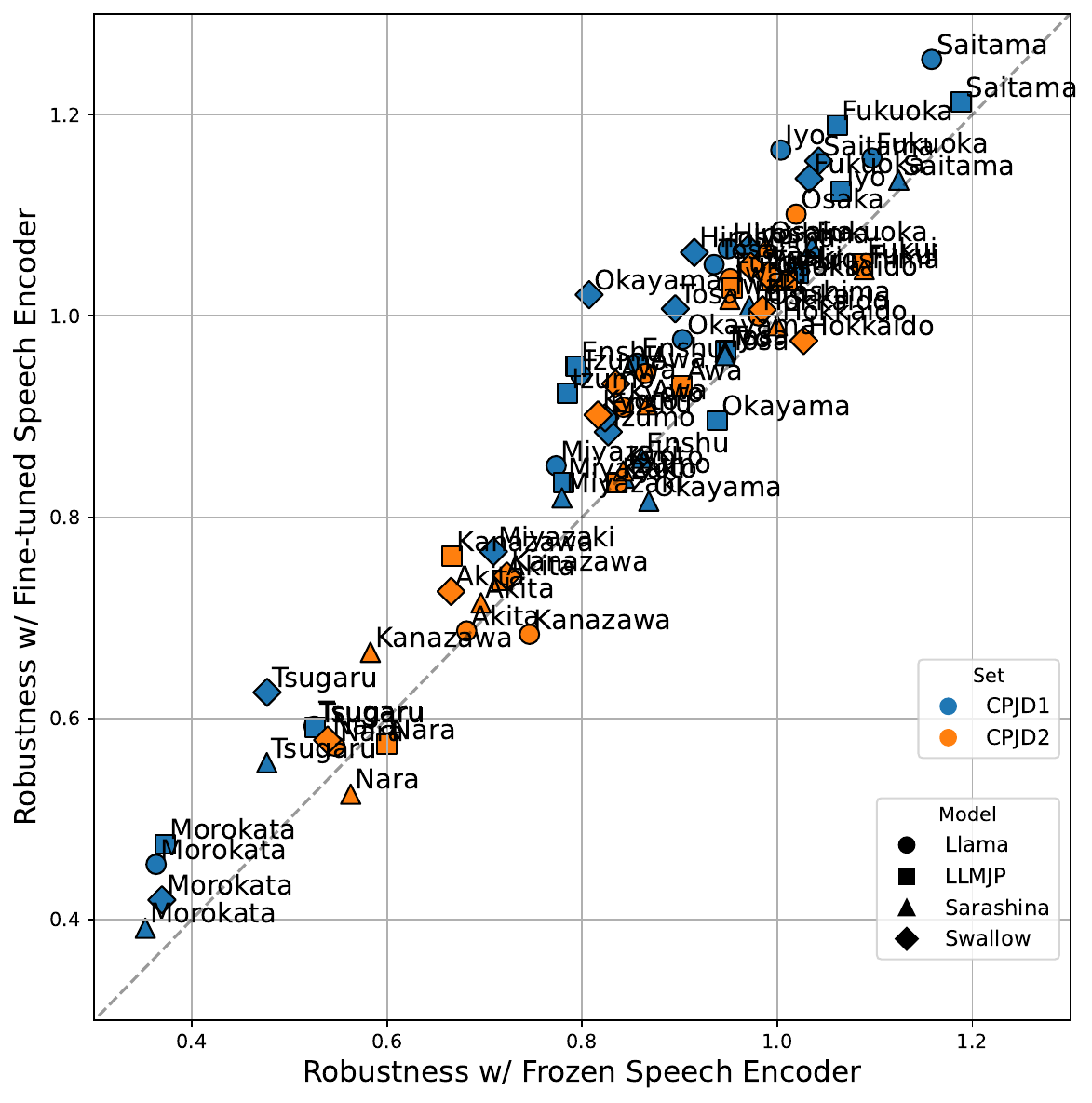}
    \vspace{-5pt}
  \caption{Robustness comparison between audio models with frozen and fine-tuned speech encoders (BLEU-based).
  The x-axis and y-axis represent dialectal robustness scores from models with frozen and fine-tuned speech encoders, respectively. } \label{fig:result005}
    \vspace{-10pt}
\end{figure}

Fig.~\ref{fig:result002} presents a comparison of dialectal robustness between text-based and audio-based models across different dialects\footnote{While robustness scores for text-based models rarely exceed 1, several audio-based models show values significantly greater than 1. This is likely an artifact of the standard Japanese input being synthesized using a TTS system.}.
Most dialects fall below the diagonal, indicating that dialectal robustness tends to decline when moving from text-based to speech-based models.
Despite this decline, a strong correlation is observed between the two model types, with Pearson correlation coefficients of 0.848 for BLEU and 0.910 for BLEURT.
This finding suggests that a model’s robustness to dialectal input in text form is a strong predictor of its robustness in spoken form.
On the other hand, the lower robustness of the SLM suggests that acoustic features such as intonation, which are not captured by the base LLM, may negatively affect the model's ability to handle dialectal speech.
Dialects such as Tsugaru, Morikata, and Akita exhibit particularly low robustness even in text-based settings.
This may be attributed to the frequent use of expressions and vocabulary that are uncommon in standard Japanese.
Although BLEU and BLEURT exhibit similar overall trends, BLEURT tends to produce scores with lower variance and smaller differences in robustness across conditions.
For clarity and due to space limitations, we therefore report only BLEU in the following sections.

\subsection{RQ2: Can dialectal training improve robustness to dialectal input?}
Fig.~\ref{fig:result004} shows dialectal robustness for models trained with and without CPJD data.
Most points lie above the diagonal, indicating that incorporating CPJD into training generally improves robustness to dialectal input.
This trend holds across both model architectures and dialect regions, although the degree of improvement varies.
These results suggest that training the adapter with dialectal data enables the model to learn alignments between dialectal speech and semantically or lexically related words.
This indicates that the model becomes better able to handle dialectal lexical variation.

It should be noted that BLEU scores for standard Japanese also shift depending on whether CPJD is included in training. 
This is because robustness is computed as a ratio, lower scores for standard Japanese can lead to an overestimation of robustness.
The scores vary within a range of approximately $-0.02$ to $+0.02$, but on average, using CPJD tends to improve performance.

\subsection{RQ3: Does fine-tuning the speech encoder improve robustness to dialectal input?}
Fig.~\ref{fig:result005} illustrates the impact of speech encoder fine-tuning on dialectal robustness.
As observed in RQ2, most points lie above the diagonal, indicating that fine-tuning the speech encoder further enhances robustness to dialectal variation.
This improvement is especially pronounced for dialects such as Morokata and Tsugaru, which initially exhibited low robustness.
These results indicate that the speech encoder, namely Whisper, is not sufficiently capable of handling Japanese dialectal speech.
To achieve robustness to dialectal variation, it is likely necessary to further train the speech encoder on dialectal data.

\section{Conclusion}
This study evaluated the dialectal robustness of LLMs and SLMs using translation tasks.
We introduced a robustness-based metric to enable fair comparisons between standard and dialectal inputs.
Our experiments demonstrated that
(1) SLMs partially inherit the dialectal capabilities of their base LLMs,
(2) training on dialectal data substantially improves robustness, and
(3) fine-tuning the speech encoder also further enhances robustness to dialectal variation.
These findings highlight the importance of dialectal supervision and targeted adaptation for building more inclusive language technologies.

\clearpage
\bibliographystyle{IEEEtran}
\bibliography{references}

@inproceedings{reazonspeech,
  title="{ReazonSpeech: A free and massive corpus for Japanese ASR}",
  author={Yin, Yue and Mori, Daijiro and Fujimoto, Seiji},
  booktitle={Proceedings of the 29th Annual Meeting of the Association for Natural Language Processing (Domestic Conference)},
  pages = {1134--1139},
  year={2023}
}

@inproceedings{shimizu-etal-2023-towards,
    title = "{Towards Speech Dialogue Translation Mediating Speakers of Different Languages}",
    author = "Shimizu, Shuichiro  and
      Chu, Chenhui  and
      Li, Sheng  and
      Kurohashi, Sadao",
    booktitle = "Findings of the Association for Computational Linguistics: ACL 2023",
    year = "2023",
    pages = "1122--1134",
}

@inproceedings{Papineni-2002-bleu,
  author = {Papineni, Kishore and Roukos, Salim and Ward, Todd and Zhu, Wei-Jing},
  title = {{BLEU: a method for automatic evaluation of machine translation}},
  year = {2002},
  booktitle = {Proceedings of the 40th Annual Meeting on Association for Computational Linguistics},
  pages = {311–318},
}

@inproceedings{takamichi-saruwatari-2018-cpjd,
    title = {{CPJD Corpus: Crowdsourced Parallel Speech Corpus of Japanese Dialects}},
    author = "Takamichi, Shinnosuke  and
      Saruwatari, Hiroshi",
    booktitle = "Proceedings of the Eleventh International Conference on Language Resources and Evaluation ({LREC} 2018)",
    year = "2018",
}

@inproceedings{sellam-etal-2020-bleurt,
    title = {{BLEURT: Learning Robust Metrics for Text Generation}},
    author = "Sellam, Thibault  and
      Das, Dipanjan  and
      Parikh, Ankur",
    booktitle = "Proceedings of the 58th Annual Meeting of the Association for Computational Linguistics",
    year = "2020",
    pages = "7881--7892",
}

@inproceedings{hu2022lora,
  title={{LoRA: Low-Rank Adaptation of Large Language Models}},
  author={Edward J Hu and Yelong Shen and Phillip Wallis and Zeyuan Allen-Zhu and Yuanzhi Li and Shean Wang and Lu Wang and Weizhu Chen},
  booktitle={Proceedings of the International Conference on Learning Representations},
  year={2022},
}

@inproceedings{Fujii:COLM2024,
  title={Continual Pre-Training for Cross-Lingual LLM Adaptation:
Enhancing Japanese Language Capabilities},
  author={Kazuki Fujii and Taishi Nakamura and Mengsay Loem and Hiroki
Iida and Masanari Ohi and Kakeru Hattori and Hirai Shota and Sakae
Mizuki and Rio Yokota and Naoaki Okazaki},
  booktitle="Proceedings of the First Conference on Language Modeling",
  year="2024",
}

@inproceedings{Okazaki:COLM2024,
  title={{Building a Large Japanese Web Corpus for Large Language Models}},
  author={Naoaki Okazaki and Kakeru Hattori and Hirai Shota and Hiroki
Iida and Masanari Ohi and Kazuki Fujii and Taishi Nakamura and Mengsay
Loem and Rio Yokota and Sakae Mizuki},
  booktitle="Proceedings of the First Conference on Language Modeling",
  year="2024",
}

@inproceedings{hono-etal-2024-integrating,
    title = {{Integrating Pre-Trained Speech and Language Models for End-to-End Speech Recognition}},
    author = "Hono, Yukiya  and
      Mitsuda, Koh  and
      Zhao, Tianyu  and
      Mitsui, Kentaro  and
      Wakatsuki, Toshiaki  and
      Sawada, Kei",
    booktitle = "Findings of the Association for Computational Linguistics: ACL 2024",
    year = "2024",
    pages = "13289--13305",
}

@inproceedings{ondrejova-suppa-2024-llms,
    title = {{Can LLMs Handle Low-Resource Dialects? A Case Study on Translation and Common Sense Reasoning in {\v{S}}ari{\v{s}}}},
    author = "Ondrejov{\'a}, Vikt{\'o}ria  and
      {\v{S}}uppa, Marek",
    booktitle = "Proceedings of the Eleventh Workshop on NLP for Similar Languages, Varieties, and Dialects (VarDial 2024)",
    year = "2024",
    pages = "130--139",
}

@inproceedings{fathullah2024,
  author={Fathullah, Yassir and Wu, Chunyang and Lakomkin, Egor and Jia, Junteng and Shangguan, Yuan and Li, Ke and Guo, Jinxi and Xiong, Wenhan and Mahadeokar, Jay and Kalinli, Ozlem and Fuegen, Christian and Seltzer, Mike},
  booktitle={Proceedings of the IEEE International Conference on Acoustics, Speech and Signal Processing}, 
  title={{Prompting Large Language Models with Speech Recognition Abilities}}, 
  year={2024},
  pages={13351-13355},
}

@INPROCEEDINGS{Lakomkin2024,
  author={Lakomkin, Egor and Wu, Chunyang and Fathullah, Yassir and Kalinli, Ozlem and Seltzer, Michael L. and Fuegen, Christian},
  booktitle={Proceedings of the IEEE International Conference on Acoustics, Speech and Signal Processing (ICASSP)}, 
  title={{End-to-End Speech Recognition Contextualization with Large Language Models}}, 
  year={2024},
  pages={12406-12410},
}

@inproceedings{kuparinen-etal-2023-dialect,
    title = {{Dialect-to-Standard Normalization: A Large-Scale Multilingual Evaluation}},
    author = "Kuparinen, Olli  and
      Mileti{\'c}, Aleksandra  and
      Scherrer, Yves",
    booktitle = "Findings of the Association for Computational Linguistics: EMNLP 2023",
    year = "2023",
    pages = "13814--13828",
}

@inproceedings{miwa2023,
  author = {Miwa, Shogo and Kai, Atsuhiko},
  year = {2023},
  pages = {4928-4932},
  title = {{Dialect Speech Recognition Modeling using Corpus of Japanese Dialects and Self-Supervised Learning-based Model XLSR}},
  booktitle={Proceedings of Interspeech 2023},
}

@inproceedings{abe-etal-2018-multi,
    title = {{Multi-dialect Neural Machine Translation and Dialectometry}},
    author = "Abe, Kaori  and
      Matsubayashi, Yuichiroh  and
      Okazaki, Naoaki  and
      Inui, Kentaro",
    booktitle = "Proceedings of the 32nd Pacific Asia Conference on Language, Information and Computation",
    year = "2018",
}

@inproceedings{paonessa-etal-2023-dialect,
    title = {{Dialect Transfer for {S}wiss {G}erman Speech Translation}},
    author = {Paonessa, Claudio  and
      Schraner, Yanick  and
      Deriu, Jan  and
      H{\"u}rlimann, Manuela  and
      Vogel, Manfred  and
      Cieliebak, Mark},
    booktitle = "Findings of the Association for Computational Linguistics: EMNLP 2023",
    year = "2023",
    pages = "15240--15254",
}

@inproceedings {Samin2024,
  author = { Sadat Samin, Md. Nazmus and Ibn Ahad, Jawad and Medha, Tanjila Ahmed and Rahman, Fuad and Amin, Mohammad Ruhul and Mohammed, Nabeel and Rahman, Shafin },
  booktitle = { 2024 IEEE International Conference on Big Data (BigData) },
  title = {{ BanglaDialecto: An End-to-End AI-Powered Regional Speech Standardization }},
  year = {2024},
  pages = {1635-1644},
}

@inproceedings{xu-etal-2018-building,
    title = {{Building Parallel Monolingual Gan Chinese Dialects Corpus}},
    author = "Xu, Fan  and
      Wang, Mingwen  and
      Li, Maoxi",
    booktitle = "Proceedings of the Eleventh International Conference on Language Resources and Evaluation ({LREC} 2018)",
    year = "2018",
}

@article{rubenstein2023audiopalm,
      title={{AudioPaLM: A Large Language Model That Can Speak and Listen}}, 
      author={Paul K. Rubenstein and Chulayuth Asawaroengchai and Duc Dung Nguyen and Ankur Bapna and Zalán Borsos and others},
      year={2023},
      journal={arXiv:2306.12925},
}

@article{chu2023qwenaudio,
      title={{Qwen-Audio: Advancing Universal Audio Understanding via Unified Large-Scale Audio-Language Models}}, 
      author={Yunfei Chu and Jin Xu and Xiaohuan Zhou and Qian Yang and Shiliang Zhang and Zhijie Yan and Chang Zhou and Jingren Zhou},
      year={2023},
      journal={arXiv:2311.07919},
}

@article{chu2024qwen2audiotechnicalreport,
    title={{Qwen2-Audio Technical Report}}, 
    author={Yunfei Chu and Jin Xu and Qian Yang and Haojie Wei and Xipin Wei and Zhifang Guo and Yichong Leng and Yuanjun Lv and Jinzheng He and Junyang Lin and Chang Zhou and Jingren Zhou},
    year={2024},
    journal={arXiv:2407.10759},
}

@article{wang2020covost,
    title={{CoVoST 2: A Massively Multilingual Speech-to-Text Translation Corpus}},
    author={Changhan Wang and Anne Wu and Juan Pino},
    year={2020},
    journal={arXiv:2007.10310},
}

@article{radford2022whisper,
  author = {Radford, Alec and Kim, Jong Wook and Xu, Tao and Brockman, Greg and McLeavey, Christine and Sutskever, Ilya},
  title = {{Robust Speech Recognition via Large-Scale Weak Supervision}},
  year = {2022},
  journal={arXiv:2212.04356},
}

@article{grattafiori2024llama3herdmodels,
  title={{The Llama 3 Herd of Models}}, 
  author={Aaron Grattafiori and Abhimanyu Dubey and Abhinav Jauhri and Abhinav Pandey and Abhishek Kadian and others},
  year={2024},
  journal={arXiv:2407.21783},
}

@article{qwen2025qwen25technicalreport,
  title={{Qwen2.5 Technical Report}}, 
  author={Qwen and : and An Yang and Baosong Yang and Beichen Zhang and Binyuan Hui and Bo Zheng and Bowen Yu and Chengyuan Li and Dayiheng Liu and Fei Huang and Haoran Wei and others},
  year={2025},
  journal={arXiv:2412.15115},
}

@article{deepseekai2024deepseekv3technicalreport,
  title={{DeepSeek-V3 Technical Report}}, 
  author={DeepSeek-AI and Aixin Liu and Bei Feng and Bing Xue and Bingxuan Wang and Bochao Wu and Chengda Lu and Chenggang Zhao and Chengqi Deng and Chenyu Zhang and others},
  year={2024},
  journal={arXiv:2412.19437},
}

@article{openai2024gpt4technicalreport,
  title={{GPT-4 Technical Report}}, 
  author={OpenAI and Josh Achiam and Steven Adler and Sandhini Agarwal and Lama Ahmad and Ilge Akkaya and Florencia Leoni Aleman and Diogo Almeida and Janko Altenschmidt and others},
  year={2024},
  journal={arXiv:2303.08774},
}

@article{oanil2023palm2technicalreport,
  title={{PaLM 2 Technical Report}}, 
  author={Rohan Anil and Andrew M. Dai and Orhan Firat and Melvin Johnson and Dmitry Lepikhin and Alexandre Passos and Siamak Shakeri and Emanuel Taropa and Paige Bailey and Zhifeng Chen and others},
  year={2023},
  journal={arXiv:2305.10403},
}

@article{lin2025languagegapsevaluatingdialect,
      title={{One Language, Many Gaps: Evaluating Dialect Fairness and Robustness of Large Language Models in Reasoning Tasks}}, 
      author={Fangru Lin and Shaoguang Mao and Emanuele La Malfa and Valentin Hofmann and Adrian de Wynter and Xun Wang and Si-Qing Chen and Michael Wooldridge and Janet B. Pierrehumbert and Furu Wei},
      year={2025},
      eprint={2410.11005},
      journal={arXiv:2410.11005},
}

@article{limkonchotiwat2025assessingthaidialectperformance,
      title={{Assessing Thai Dialect Performance in LLMs with Automatic Benchmarks and Human Evaluation}}, 
      author={Peerat Limkonchotiwat and Kanruethai Masuk and Surapon Nonesung and Chalermpun Mai-On and Sarana Nutanong and Wuttikorn Ponwitayarat and Potsawee Manakul},
      year={2025},
      eprint={2504.05898},
      journal={arXiv:2504.05898},
}

\end{document}